\documentclass[12pt]{article}
\usepackage{axodraw}

\catcode`@=12
\begin{document}
\thispagestyle{empty}
\begin{flushright} 
UCRHEP-T349\\ 
November 2002\
\end{flushright}
\vspace{0.5in}
\begin{center}
{\LARGE	\bf Plato's Fire and the Neutrino Mass Matrix\\}
\vspace{1.2in}
{\bf Ernest Ma\\}
\vspace{0.2in}
{\sl Physics Department, University of California, Riverside, 
California 92521\\}
\vspace{1.2in}
\end{center}
\begin{abstract}\
With the accumulation of many years of solar and atmospheric neutrino 
oscillation data, the approximate form of the $3 \times 3$ neutrino mixing 
matrix is now known.  The theoretical challenge is to understand where this 
mixing matrix comes from.  Recently, a remarkable fact was discovered that 
for a specific pattern of the neutrino mass matrix at a high scale, any 
flavor-changing radiative correction will automatically lead to the desired 
mixing matrix.  It was also discovered that the required specific pattern at 
the high scale can be maintained by the non-Abelian discrete symmetry $A_4$ 
which is also the symmetry group of the regular tetrahedron, one of five 
perfect geometric solids known to Plato who associated it with the element 
``fire''.  I discuss this recent development and add to it a new and very 
simple mechanism for the implementation of the flavor-changing radiative 
correction.
\end{abstract}
\vspace{0.2in}
\noindent ------------------

\noindent To appear as a Brief Review in Modern Physics Letters A
\newpage
\baselineskip 24pt

\section{Introduction}

With the recent addition of the SNO (Sudbury Neutrino Observatory) 
neutral-current data \cite{sno}, the overall picture of solar neutrino 
oscillations \cite{sol} is becoming quite clear.  Together with the 
well-established atmospheric neutrino data \cite{atm}, the $3 \times 3$ 
neutrino mixing matrix is now determined to a very good first approximation by
\begin{equation}
\pmatrix {\nu_e \cr \nu_\mu \cr \nu_\tau} = \pmatrix {\cos \theta & -\sin 
\theta & 0 \cr \sin \theta/\sqrt 2 & \cos \theta/\sqrt 2 & -1/\sqrt 2 \cr 
\sin \theta/\sqrt 2 & \cos \theta/\sqrt 2 & 1/\sqrt 2} \pmatrix {\nu_1 \cr 
\nu_2 \cr \nu_3},
\end{equation}
where $\nu_{1,2,3}$ are assumed to be Majorana neutrino mass eigenstates.  
In the above, $\sin^2 2 \theta_{atm} = 1$ is already assumed and $\theta$ 
is the solar mixing angle which is now known to be large but not maximal 
\cite{many}, i.e. $\tan^2 \theta \simeq 0.4$.  The $U_{e3}$ entry has been 
assumed zero but it is only required experimentally to be small \cite{react}, 
i.e. $|U_{e3}| < 0.16$. 

Denoting the masses of $\nu_{1,2,3}$ as $m_{1,2,3}$, the solar neutrino data 
\cite{sno,sol} require that $m_2^2 > m_1^2$ with $\theta < \pi/4$, and in 
the case of the favored large-mixing-angle solution \cite{many},
\begin{equation}
\Delta m^2_{sol} = m_2^2 - m_1^2 \simeq 5 \times 10^{-5}~{\rm eV}^2.
\end{equation}
The atmospheric neutrino data \cite{atm} require
\begin{equation}
|m_3^2 - m_{1,2}^2| \simeq 2.5 \times 10^{-3}~{\rm eV}^2,
\end{equation}
without deciding whether $m_3^2 > m_{1,2}^2$ or $m_3^2 < m_{1,2}^2$.

The big question now is what the neutrino mass matrix itself should look 
like, in order that Eq.~(1) be obtained.  Since neutrino oscillations 
measure only the differences of the squares of the mass eigenvalues, it 
is obvious that the answer will not be unique \cite{ma02}.  On the other 
hand, a pattern supported by an underlying symmetry would be better motivated 
than an $ad~hoc$ hypothesis.  In this Brief Review, a recent interesting 
development in this direction is presented, with details for a more general 
readership as well as something entirely new.

In Sec.~2, it is shown how Eq.~(1) may be automatically obtained from $any$ 
flavor-changing radiative correction of a particular neutrino mass matrix 
at a high scale.  This of course requires physics beyond the Standard Model. 
In Sec.~3, the symmetry which allows us to have that special neutrino mass 
matrix at the high scale is identified as Plato's ``fire'', i.e. the 
non-Abelian discrete symmetry $A_4$.  In Sec.~4, it is shown how the 
irreducible representations of $A_4$ are just right to allow for three 
arbitrary charged-lepton masses while maintaining three degenerate neutrino 
masses at the high scale.  In Sec.~5, a new and very simple mechanism for 
flavor-changing radiative corrections is proposed, which gives realistic 
values for the neutrino mass differences of Eqs.~(2) and (3).  In Sec.~6, 
there are some concluding remarks.

\section{Getting the Right Neutrino Mixing Matrix With Almost Nothing}

Given the particle content of the Standard Model, lepton masses come from
\begin{equation}
{\cal L}_{int} = f_{ij} \overline {(\nu, e)}_{iL} e_{jR} \pmatrix {\phi^+ \cr 
\phi^0} + \lambda_{ij} (\nu_i \phi^0 - e_i \phi^+)(\nu_j \phi^0 - e_j \phi^+) 
+ H.c.,
\end{equation}
where the second term is the effective dimensional-five operator \cite{wein79} 
for Majorana neutrino masses.  Note that $\lambda_{ij}$ has the dimension of 
inverse mass, hence any such neutrino mass must be proportional to the square 
of $v = \langle \phi^0 \rangle$ divided by a large mass, i.e. ``seesaw'' in 
character whatever its origin \cite{ma98}.  At some high scale, let
\begin{equation}
f_{ij} = U_L^\dagger \pmatrix {f_e & 0 & 0 \cr 0 & f_\mu & 0 \cr 0 & 0 & 
f_\tau} U_R,
\end{equation}
then in the $\nu_{e,\mu,\tau}$ basis, $\lambda_{ij}$ becomes $U_L^T 
\lambda_{ij} U_L$ which is of course arbitrary without further assumptions.  
Suppose however that for some reason,
\begin{equation}
U_L^T \lambda_{ij} U_L \propto \pmatrix {1 & 0 & 0 \cr 0 & 0 & 1 \cr 0 & 1 & 
0},
\end{equation}
then a remarkable thing happens when one-loop flavor-changing radiative 
corrections are added, i.e. Eq.~(1) will be automatically obtained, as shown 
below.  Note that the form of Eq.~(6) is crucial for this to hold.  If it 
were simply proportional to the identity matrix, Eq.~(1) would not be the 
result.

From the high scale to the electroweak scale, one-loop radiative corrections 
will change Eq.~(6) to
\begin{equation}
\pmatrix {1 & 0 & 0 \cr 0 & 0 & 1 \cr 0 & 1 & 0} +  R \pmatrix {1 & 0 & 0 
\cr 0 & 0 & 1 \cr 0 & 1 & 0} + \pmatrix {1 & 0 & 0 \cr 0 & 0 & 1 \cr 0 & 1 
& 0} R^T,
\end{equation}
where the radiative correction matrix is assumed to be of the most general 
form, i.e.
\begin{equation}
R = \pmatrix {r_{ee} & r_{e\mu} & r_{e\tau} \cr r_{e\mu}^* & r_{\mu\mu} & 
r_{\mu\tau} \cr r_{e\tau}^* & r_{\mu\tau}^* & r_{\tau\tau}}.
\end{equation}
Thus the observed neutrino mass matrix is given by
\begin{equation}
{\cal M}_\nu = m_0 \pmatrix {1+2r_{ee} & r_{e\tau} + r_{e\mu}^* & r_{e\mu} + 
r_{e\tau}^* \cr r_{e\mu}^* + r_{e\tau} & 2r_{\mu\tau} & 1+r_{\mu\mu}+
r_{\tau\tau} \cr r_{e\tau}^* + r_{e\mu} & 1+r_{\mu\mu}+r_{\tau\tau} & 
2r_{\mu\tau}^*}.
\end{equation}
Then using the redefinitions:
\begin{eqnarray}
&& \delta_0 \equiv r_{\mu\mu} + r_{\tau\tau} - r_{\mu\tau} - r^*_{\mu \tau}, 
\\ 
&& \delta \equiv 2r_{\mu\tau}, \\
&& \delta' \equiv r_{ee} - {1 \over 2} r_{\mu\mu} - {1 \over 2} r_{\tau\tau} 
- {1 \over 2} r_{\mu\tau} - {1 \over 2} r^*_{\mu \tau}, \\
&& \delta'' \equiv r_{e\mu}^* + r_{e\tau},
\end{eqnarray}
the neutrino mass matrix becomes
\begin{equation}
{\cal M}_\nu = m_0 \pmatrix{1+\delta_0+\delta+\delta^*+2\delta' & \delta'' & 
\delta''^* \cr \delta'' & \delta & 1+\delta_0+(\delta+\delta^*)/2 \cr 
\delta''^* & 1+\delta_0+(\delta+\delta^*)/2 & \delta^*}.
\end{equation}
Without any loss of generality, $\delta$ may be chosen real by absorbing its 
phase into $\nu_\mu$ and $\nu_\tau$ and $\delta_0$ set equal to zero by 
redefining $m_0$ and the other $\delta$'s.

In the Standard Model, there are no flavor-changing leptonic interactions, 
thus $\delta = \delta'' = 0$ and Eq.~(14) does not lead to neutrino 
oscillations at all.  However, if there is some new physics which allows 
all the $\delta$'s to be nonzero, then Eq.~(14) is exactly diagonalizable 
if $\delta''$ is real, and the result is Eq.~(1) independent of the actual 
values of $m_0$, $\delta$, $\delta'$, or $\delta''$, subject only to the 
constraint
\begin{equation}
\tan \theta = {\sqrt 2 \delta'' \over \sqrt {\delta'^2 + 2 \delta''^2} - 
\delta'}, ~~~{\rm with} ~\delta' < 0.
\end{equation}
The mass eigenvalues do depend on $m_0$ and the $\delta$'s and they are 
given by
\begin{eqnarray}
m_1 &=& m_0 (1 + 2 \delta + \delta' - \sqrt {\delta'^2 + 2 \delta''^2} ), \\ 
m_2 &=& m_0 (1 + 2 \delta + \delta' + \sqrt {\delta'^2 + 2 \delta''^2} ), \\ 
m_3 &=& -m_0.
\end{eqnarray}
Thus the relevant $\Delta m^2$ parameters for atmospheric and solar neutrino 
oscillations are respectively
\begin{equation}
(\Delta m^2)_{atm} = (\Delta m^2)_{32} = (\Delta m^2)_{31} \simeq 4 \delta 
m_0^2, ~~~ (\Delta m^2)_{sol} = (\Delta m^2)_{12} \simeq 4 \sqrt {\delta'^2 + 
2 \delta''^2} m_0^2.
\end{equation}
To obtain $U_{e3} \neq 0$, let $Im \delta'' \neq 0$ but small, then
\begin{equation}
U_{e3} \simeq {i Im \delta'' \over \sqrt 2 \delta},
\end{equation}
and the above expressions are corrected by the replacement of $\delta'$ with 
$\delta' + (Im \delta'')^2/2 \delta$, and $\delta''$ by $Re \delta''$.  This 
results in the following interesting relationship:
\begin{equation}
\left[ {(\Delta m^2)_{12} \over (\Delta m^2)_{32}} \right]^2 \simeq 
\left[ {\delta' \over \delta} + |U_{e3}|^2 \right]^2 + 2 \left[ {Re \delta'' 
\over \delta} \right]^2.
\end{equation}

It has thus been shown that with Eq.~(6) at the high scale, a desirable 
neutrino mixing matrix is automatically generated by arbitrary radiative 
corrections, including the possibility of CP violation which is predicted 
here to be $maximal$ because $U_{e3}$ is purely imaginary.

\section{Plato's Fire}

The patterns of Eqs.~(5) and (6) should be simultaneously maintained by some 
symmetry, but it looks impossible.  However, there is in fact a solution, and 
it is based on the non-Abelian discrete symmetry $A_4$ \cite{a4,bmv}. 
What is $A_4$ and why is it special?

Around the year 390 BCE, the Greek mathematician Theaetetus proved that there 
are five and only five perfect geometric solids.  The Greeks already knew 
that there are four basic elements: fire, air, water, and earth.  Plato 
could not resist matching them to the five perfect geometric solids and 
for that to work, he invented the fifth element, i.e. quintessence, which 
is supposed to hold the cosmos together.  His assignments are shown in 
Table 1.

\begin{table}[htb]
\caption{Properties of Perfect Geometric Solids}
\begin{center}
\begin{tabular}{|c|c|c|c|c|}
\hline 
solid & faces & vertices & Plato & Group \\ 
\hline
tetrahedron & 4 & 4 & fire & $A_4$ \\ 
octahedron & 8 & 6 & air & $S_4$ \\ 
icosahedron & 20 & 12 & water & $A_5$ \\ 
hexahedron & 6 & 8 & earth & $S_4$ \\ 
dodecahedron & 12 & 20 & ? & $A_5$ \\ 
\hline
\end{tabular}
\end{center}
\end{table}

The group theory of these solids was established in the early 19th century. 
Since a cube (hexahedron) can be imbedded perfectly inside an octahedron 
and the latter inside the former, they have the same symmetry group.  
The same holds for the icosahedron and dodecahedron.  The tetrahedron 
(Plato's ``fire'') is special because it is self-dual.  It has the symmetry 
group $A_4$, i.e. the finite group of the even permutation of 4 objects. 
The reason that it is special for the neutrino mass matrix is because it 
has 3 inequivalent one-dimensional irreducible representations and 1 
three-dimensional irreducible representation exactly.  Its character table 
is given below.

\begin{table}[htb]
\caption{Character Table of $A_4$}
\begin{center}
\begin{tabular}{|c|c|c|c|c|c|c|}
\hline
class & n & h & $\chi_1$ & $\chi_2$ & $\chi_3$ & $\chi_4$ \\ 
\hline
$C_1$ & 1 & 1 & 1 & 1 & 1 & 3 \\ 
$C_2$ & 4 & 3 & 1 & $\omega$ & $\omega^2$ & 0 \\ 
$C_3$ & 4 & 3 & 1 & $\omega^2$ & $\omega$ & 0 \\ 
$C_4$ & 3 & 2 & 1 & 1 & 1 & $-1$ \\ 
\hline
\end{tabular}
\end{center}
\end{table}

In the above, $n$ is the number of elements, $h$ is the order of each element, 
and
\begin{equation}
\omega = e^{2 \pi i/3}
\end{equation}
is the cube root of unity.  The group multiplication rule is
\begin{equation}
\underline {3} \times \underline {3} = \underline {1} + \underline {1}' + 
\underline {1}'' + \underline {3} + \underline {3}.
\end{equation}

\section{Details of the $A_4$ Model}

Using $A_4$, we now have the following natural assignments of quarks and 
leptons:
\begin{eqnarray}
&& (u_i,d_i)_L, ~~ (\nu_i,e_i)_L \sim \underline {3}, \\ 
&& u_{1R}, ~~ d_{1R}, ~~ e_{1R} \sim \underline {1}, \\ 
&& u_{2R}, ~~ d_{2R}, ~~ e_{2R} \sim \underline {1}', \\ 
&& u_{3R}, ~~ d_{3R}, ~~ e_{3R} \sim \underline {1}''.
\end{eqnarray}
Heavy fermion singlets are then added \cite{bmv}:
\begin{equation}
U_{iL(R)}, ~~ D_{iL(R)}, ~~ E_{iL(R)}, ~~ N_{iR} \sim \underline {3},
\end{equation}
together with the usual Higgs doublet and new heavy singlets:
\begin{equation}
(\phi^+,\phi^0) \sim \underline {1}, ~~~~ \chi^0_i \sim \underline {3}.
\end{equation}
With this structure, charged leptons acquire an effective Yukawa coupling 
matrix $\bar e_{iL} e_{jR} \phi^0$ which has 3 arbitrary eigenvalues 
(because of the 3 independent couplings to the 3 inequivalent one-dimensional 
representations) and for the case of equal vacuum expectation values of 
$\chi_i$, i.e.
\begin{equation}
\langle \chi_1 \rangle = \langle \chi_2 \rangle = \langle \chi_3 \rangle = u,
\end{equation}
the unitary transformation $U_L$ which diagonalizes $f_{ij}$ of Eq.~(5) is 
given by
\begin{equation}
U_L = {1 \over \sqrt 3} \pmatrix {1 & 1 & 1 \cr 1 & \omega & \omega^2 \cr 
1 & \omega^2 & \omega}.
\end{equation}
The corresponding $\lambda_{ij}$ of Eq.~(6) is fixed by $A_4$ to be 
proportional to the identity matrix; thus the effective neutrino 
mass operator, i.e. $\nu_i \nu_j \phi^0 \phi^0$, is proportional to
\begin{equation}
U_L^T U_L = \pmatrix {1 & 0 & 0 \cr 0 & 0 & 1 \cr 0 & 1 & 0},
\end{equation}
exactly as desired \cite{a4,bmv}.

To derive Eq.~(32), the validity of Eq.~(30) has to be proved.  This is 
naturally accomplished in the context of supersymmetry.  Let $\hat \chi_i$ 
be superfields, then its superpotential is given by
\begin{equation}
\hat W = {1 \over 2} M_\chi (\hat \chi_1 \hat \chi_1 + \hat \chi_2 \hat \chi_2 
+ \hat \chi_3 \hat \chi_3) + h_\chi \hat \chi_1 \hat \chi_2 \hat \chi_3.
\end{equation}
Note that the $h_\chi$ term is invariant under $A_4$, a property not found 
in $SU(2)$ or $SU(3)$.  The resulting scalar potential is
\begin{equation}
V = |M_\chi \chi_1 + h_\chi \chi_2 \chi_3|^2 + |M_\chi \chi_2 + h_\chi \chi_3 
\chi_1|^2 + |M_\chi \chi_3 + h_\chi \chi_1 \chi_2|^2.
\end{equation}
Thus a supersymmetric vacuum $(V=0)$ exists for
\begin{equation}
\langle \chi_1 \rangle = \langle \chi_2 \rangle = \langle \chi_3 \rangle = u 
= -M_\chi /h_\chi,
\end{equation}
proving Eq.~(30), with the important additional result that the spontaneous 
breaking of $A_4$ at the high scale $u$ does not break the supersymmetry.

\section{New Flavor-Changing Radiative Mechanism}

The original $A_4$ model \cite{a4} was conceived to be a symmetry at the 
electroweak scale, in which case the splitting of the neutrino mass 
degeneracy is put in by hand and any mixing matrix is possible.  Subsequently, 
it was proposed \cite{bmv} as a symmetry at a high scale, in which case the 
mixing matrix is determined completely by flavor-changing radiative 
corrections and the only possible result happens to be Eq.~(1) if CP is 
conserved.  This is a remarkable convergence in that Eq.~(1) is also the 
phenomenologically preferred neutrino mixing matrix based on the most recent 
data from neutrino oscillations.

We should now consider the new physics responsible for the $\delta$'s of 
Eq.~(14).  Previously \cite{bmv}, arbitrary soft supersymmetry breaking in 
the scalar sector was invoked.  It is certainly a phenomenologically viable 
scenario, but lacks theoretical motivation and is somewhat complicated.  Here 
I propose a new and much simpler mechanism, using a triplet of charged 
scalars under $A_4$, i.e. $\eta^+_i \sim \underline {3}$.  Their relevant 
contributions to the Lagrangian of this model is then
\begin{equation}
{\cal L} = f \epsilon_{ijk} (\nu_i e_j - e_i \nu_j) \eta^+_k + m_{ij}^2 
\eta^+_i \eta^-_j.
\end{equation}
Whereas the first term is invariant under $A_4$ as it should be, the second 
term is a soft term which is allowed to break $A_4$, from which the 
flavor-changing radiative corrections will be calculated.

The neutrino wave-function renormalizations are depicted in Fig.~1, where 
the indices $i,j,k$ are all different.  Let
\begin{equation}
\pmatrix {\eta_e \cr \eta_\mu \cr \eta_\tau} = \pmatrix {U_{e1} & U_{e2} & 
U_{e3} \cr U_{\mu 1} & U_{\mu 2} & U_{\mu 3} \cr U_{\tau 1} & U_{\tau 2} & 
U_{\tau 3}} \pmatrix {\eta_1 \cr \eta_2 \cr \eta_3},
\end{equation}
where $\eta_{1,2,3}$ are mass eigenstates with masses $m_{1,2,3}$.  Using 
Eqs.~(11) to (13), the $\delta$'s are then all easily calculated.  They are 
all finite and independent of an arbitrary overall mass scale as shown below.
\begin{eqnarray}
\delta &=& -{f^2 \over 4 \pi^2} \sum_{i=1}^3 U^*_{\mu i} U_{\tau i} \ln m_i^2,
\\  
{\delta}' &=& -{f^2 \over 8 \pi^2} \sum_{i=1}^3 \left( {1 \over 2} 
|U_{\mu i}-U_{\tau i}|^2 - |U_{e i}|^2 \right) \ln m_i^2, \\ 
{\delta}'' &=& -{f^2 \over 8 \pi^2} \sum_{i=1}^3 ( U^*_{\mu i} U_{e i} + 
U_{\tau i} U^*_{e i} ) \ln m_i^2.
\end{eqnarray}

\begin{figure}
\begin{center}
\begin{picture}(360,80)(0,0)
\ArrowLine(0,10)(40,10)
\ArrowLine(120,10)(40,10)
\ArrowLine(120,10)(160,10)
\ArrowLine(200,10)(240,10)
\ArrowLine(320,10)(240,10)
\ArrowLine(320,10)(360,10)
\DashArrowArc(80,10)(40,90,180){4}
\DashArrowArc(80,10)(40,0,90){4}
\DashArrowArc(280,10)(40,0,180){4}

\Text(20,0)[]{$\nu_i$}
\Text(80,0)[]{$e_k$}
\Text(140,0)[]{$\nu_j$}
\Text(220,0)[]{$\nu_i$}
\Text(280,0)[]{$e_{j,k}$}
\Text(340,0)[]{$\nu_i$}
\Text(80,50)[]{$\times$}
\Text(40,40)[]{$\eta_j^+$}
\Text(123,40)[]{$\eta_i^+$}
\Text(280,62)[]{$\eta^+_{k,j}$}
\end{picture}
\end{center}
\caption{Neutrino wave-function renormalizations.}
\end{figure}
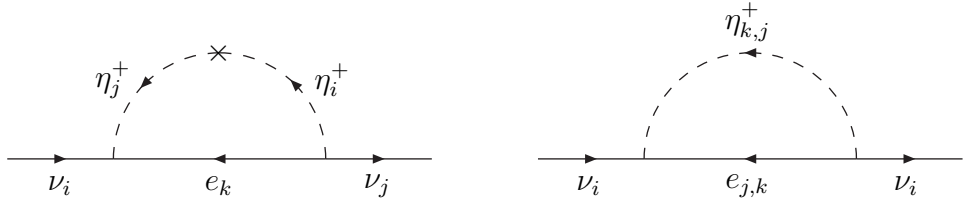

As an example, consider
\begin{equation}
U = \pmatrix {1/\sqrt 2 & -1/\sqrt 2 & 0 \cr c/\sqrt 2 & c/\sqrt 2 & s \cr 
-s/\sqrt 2 & -s/\sqrt 2 & c},
\end{equation}
then
\begin{eqnarray}
\delta &=& {f^2 s c \over 8 \pi^2} \left( \ln {m_1^2 \over m_3^2} + \ln 
{m_2^2 \over m_3^2} \right), \\ 
\delta' &=& {f^2 (1-2sc) \over 32 \pi^2} \left( \ln {m_1^2 \over m_3^2} + \ln 
{m_2^2 \over m_3^2} \right), \\ 
\delta'' &=& {f^2 (s-c) \over 16 \pi^2} \ln {m_1^2 \over m_2^2}.
\end{eqnarray}
To obtain Eqs.~(2), (3), and $\tan^2 \theta \simeq 0.4$, using Eqs.~(15) and 
(19), a possible solution is
\begin{equation}
s = 0.638, ~~ c=0.770, ~~ {m_3^2 \over m_1^2} = 1.728, ~~ {m_3^2 \over m_2^2} 
= 1.573, ~~ f^2 m_0^2 = 0.1 ~{\rm eV}^2,
\end{equation}
where $m_0$ is the approximate common mass of all 3 neutrinos, as measured 
in neutrinoless double beta decay \cite{klapdor}.

To determine the absolute values of $m_{1,2,3}$, the most stringent condition 
comes from the flavor-changing radiative decay $\mu \to e \gamma$.  Its 
amplitude is given here by
\begin{equation}
{\cal A} = {e f^2 c m_\mu \over 192 \pi^2} \left( {1 \over m_1^2} - 
{1 \over m_2^2} \right) \epsilon^\lambda q^\nu \bar e \sigma_{\lambda \nu} 
(1+\gamma_5) \mu.
\end{equation}
The resulting branching fraction is then
\begin{equation}
B (\mu \to e \gamma) = 4.24 \times 10^{-10} f^4 \left( {1~{\rm TeV} \over m_1} 
\right)^4.
\end{equation}
Using the present experimental upper bound of $1.2 \times 10^{-11}$, the 
constraint
\begin{equation}
m_1/f > 2.44 ~{\rm TeV}
\end{equation}
is obtained.

\section{Concluding Remarks}

In conclusion, recent experimental progress on neutrino oscillations points 
to a neutrino mixing matrix which can be derived in a remarkable way through 
radiative corrections of an underlying $A_4$ symmetry at some high scale. 
This scheme \cite{bmv} automatically leads to $\sin^2 2 \theta_{atm} = 1$ 
and a large (but not maximal) solar mixing angle.  A new and very simple 
radiative mechanism using a triplet of heavy charged scalars (at the TeV 
scale) is proposed which leads to realistic values of the neutrino-oscillation 
parameters.  To the extent that the Yukawa coupling $f$ should not be too big, 
the value of $m_0$ measured in neutrinoless double beta decay should not be 
much less than its current upper bound of about 0.4 eV.

This work was supported in part by the U.~S.~Department of Energy
under Grant No.~DE-FG03-94ER40837.

%\newpage
\bibliographystyle{unsrt}

\end{document}